# Ontology-oriented e-gov services retrieval

Hassania Ouchetto[1], Ouail Ouchetto[2] and Ounsa Roudiès[1]

[1] Computer Science Department, Mohammadia School of Ingineers (EMI), Mohammed V[th] University-Agdal
10000 Rabat, Morocco
ouchetto@affaires-generales.gov.ma
roudies@emi.ac.ma

[2] University Aïn Chock Hassan II, Casablanca
20000 Casablanca, Morocco
ouail.ouchetto@gmail.com

**Abstract**
The semantic e-government is a new application field accompanying the development of semantic web where the ontologies have become a fertile field of investigation. This is due firstly to both the complexity and the size of e-government systems and secondly to the importance of the issues. However, permitting easy and personalized access to e-government services has become, at this juncture, an arduous and not spontaneous process. Indeed, the provided e-gov services to the user represent a critical contact point between administrations and users. The encountered problems in the e-gov services retrieving process are: the absence of an integrated one-stop government, the difficulty of localizing the services' sources, the lack of mastery of search terms and the deficiency of multilingualism of the online services. In order to solve these problems, to facilitate access to e-gov services and to satisfy the needs of potential users, we propose an original approach to this issue. This approach incorporates a semantic layer as a crucial element in the retrieving process. It consists in implementing a personalized search system that integrates ontology of the e-gov domain in this process.

## 1. Introduction

Nowadays, Morocco is witnessing a strong hankering for e-government services (online administrative services) given the facilities they provide for organizations and citizens alike. In actual fact, the digital and information technologies era introduced a major change in the Moroccan society, revolutionizing the behaviour of economic players as well as that of individuals in their daily life.

In addition, e-government modifies interactions between public powers on the one hand and, on the other hand, between organizations and citizens [1] [2]. This is effected with great:
1. Rapidity: in so far as the user does not need to be physically present at an administration office since s/he can carry out such transaction online. Moreover, the new data collection and exchange procedure is likely to limit contacts between citizens and organizations with administrations.
2. Accessibility: given the possibility of accessing administrative offices electronically around the clock and during all weekdays wherever the user is.
3. Transparency: given that e-gov facilitates transactions among public powers on the one hand and, on the other hand, among citizens and organizations. It tends to involve users in the government decision making processes.
4. Cost-effectiveness, as e-gov permits significant reductions in favour of taxpayers and guarantees quality of the offered services.

In other words, the main objective of e-government is to develop technological solutions in order to support the interaction between citizens, organizations and public institutions, and will improve their participation in public and social life [3].

Therefore, e-gov is by no means limited to a mere publishing of information on public administration's websites. Rather, it involves a profound reconsideration of the structure and functions of administrations. As a matter of fact, administrative procedures such as electronic data collection, processing and exchange within or among administrations, need to be adapted to the electronic public services provided to satisfy the citizens and the organizations requirements.

The action plan developed by the Moroccan e-gov commission, responsible for governmental projects, has targeted a considerable number of undertaking. These projects are so diverse and are organized either thematically (health, education, finance, etc.), or by user (citizens, organizations, local communities). They projects fall into two major categories. The first category is related to particular sectors such as e-justice, e-finance, DAMANCOM (CNSS)—a social security organism—, the National Agency for Land Conservation, the Cadastre and Cartography, e-health, e-consulate…etc. The second category concerns transverse projects such as the National Portal, the e-Wilaya and the public services.

However, at this juncture, easy and personalized access to e-government services has become an arduous process. In fact, the e-gov services represent a critical contact point between administrations and users. The problems encountered in the e-gov services retrieving process are as follows:

1. User disorientation: each administration has its own independent solution. It cannot provide an integrated service in one-stop government that offers all the services. Therefore, the search for a specific service is carried out through the portal of the administration that owns the service in question.
2. Difficulty of localizing the sources of services: knowing the internet addresses of service portals of every administration is not always an easy undertaking. It is an operation that requires considerable effort which consists in a thorough research on the web.
3. Lack of mastery of search terms: the used terms in search for a desired service do not necessarily correspond to those found in the vocabulary controlled both by the service suppliers and experts in the field. This is due to the absence of a semantic layer in the e-gov systems.
4. Multilingualism: The services proposed by the different Moroccan administrations are generally published either in French or in Arabic. Consequently, a user who launches his search in Arabic that is actually published in French, and vice versa, runs the risk of failing to find what s/he is searching for.

In order to solve these problems, to satisfy the needs of potential users and to facilitate access to e-gov services, we propose an original approach to this issue. Our approach incorporates a semantic layer as a crucial element in the retrieving process. This consists in implementing a personalized search system that integrates ontology of the e-gov domain in this process. Our argument hinges on the following principle: when the user sends a query, the search process is automatically launched and the ontology enriches the query by new terms. This ontology is composed of several sectoral ontologies: health, tourism, customs, finance…etc.

The rest of this paper is organized as follows: the second section presents some related work with particular focus on the approaches that are based on the ontology approach. In doing so, we situate our work in the context of the e-gov semantic. The third section introduces the conceptualisation of our e-gov ontology which involves several sectoral ontologies. The fourth section is devoted to the development of our ontology as well as the approach adopted in this process. The fifth section is dedicated to the integration of this ontology in our search system. The conclusion of this paper addresses some possible follow-up perspectives.

## 2. Background and related work

In the literature, several studies have been undertaken in the field of e-government ontologies. The latter have aimed at capturing the diversity characterising a specific domain. According to Swartout and al. [4], an ontology is a group of hierarchically structured concepts designed to describe a domain and can serve as a framework for a knowledge base. The idea is that an ontology provides a skeletal structure for a knowledge base [5] [6].

In this context, the efforts invested by different research communities have given birth to the semantic e-government, which made so many projects see light. Table 1 presents some of the major e-government projects that are based on the ontological approach. It should be pointed out that the order of these projects in this table is immaterial.

The TERREGOV platform, which was destined to European employees, has two objectives. The first objective is to provide a simple terminology that allow indexing the textual documents and constitute the data base of the local agency. The second objective is to facilitate the discovery of web services that are published in directories such as UDDI [7] [8] [9].

The eGTPM was intended to help e-gov projet collaborators to exchange resources, approaches and solutions to problems met in implementing these projects. The specific objectives of eGTPM are (i) improving the performance management of public organizations in terms of efficiency, transparency and quality of rendered services and (ii) enhancing the reputation of public administrations through a good organisational management and a better internal technical management of the processes [10].

The ontoGov represents a platform that facilitates the composition, the configuration and evolution of e-gov services. Its main role is, on the one hand, to offer public administrations a global view of service configuration models, and on the other hand to improve given services to users. These processes are based on an advanced knowledge of these services [11].

The QeGS ontology is a multi-perspective evaluation of e-gov service quality. It consists of a set of high level concepts and relations between these concepts that describe the notion of service quality. This ontology forms a reusable basis for the construction of future public service systems evaluation based on the ontological approach [12].

The state project of Schleswig-Holstein is an application of the Access-eGOV project. Its principle aim is to test the integration of the existing services in a semantic web environment. Put differently, the objective is to integrate e-gov services in a semantic base, while keeping data and their maintenance in the municipality's existing systems [13]. The smartGov is, likewise, a knowledge-oriented platform that permits helping employees with their online transactions [14].

The existing e-gov projects, based on ontological approach, deal with different issues. The use of a semantic layer in these projects is varied, ranging from the indexing of e-government documents to the quality of services, through the configuration, the integration and the service migration. However, we remark that none of these projects has dealt with the integration of e-government ontology in the process of the services retrieving. This is due to the development of e-gov systems that are basically integrated into developed countries. Contrary to such developing countries as Morocco, e-government initiatives are developed in parallel, which has really complicated the search for adequate e-government services. It is noteworthy that the construction of e-government ontology is based on governmental knowledge models [15]. The latter is strongly related to laws and regulations of given governments, which differs from one country to another. Consequently, the reuse of another country's existing ontology seems impossible.

In order to satisfy this need, we have implemented an original approach for searching e-government services. This approach consists in the elaboration of a personalized search system. This system is founded, on the one hand, on the integration of a semantic layer describing a set of concepts that are managed by the Moroccan government. The approach consists, on the other hand, in taking into consideration the user profile. In the present article, we are particularly interested in introducing a sectoral ontology that is devoted to the custom's domain and its integration in the process of personalized retrieving of e-government services.

Table 1 : e-gov projects based on ontological approaches

| Project | Domain | Objectives |
|---|---|---|
| TERREGOV | Semantic e-gov | Facilitates the discovery of web services published in directories such as UDDI. |
| eGTMP | Semantic e-gov | e-gov management projects |
| OntoGov | Semantic e-gov | A platform proposed for the composition, the configuration and the evolution of e-gov services. |
| SmartGov | Semantic e-gov | Knowledge-oriented platform facilitating online transactions for employees. |
| QeGS | Semantic e-gov | Multi-perspective evaluation ontology for the quality of e-gov services. |
| Schleswig-Holstein state | Semantic e-gov | The migration of a website to a semantic web environment. |

In the following section, we present the design of our e-government ontology that is composed of a number of sectoral ontologies.

## 3. Design of the ontology

Because the terminology relating to the domain of e-government is both vast and varied, we propose to divide our e-government domain ontology into sub-ontologies for particular domains such as tourism, customs, finance, etc. (cf. fig. 1.).

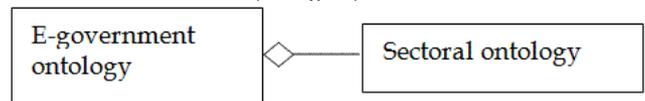

Fig. 1 Sectoral ontology

Sectoral ontologies are constructed in an independent fashion. Then, they are reused for the construction of mother ontology of the e-government domain. The figure 2 introduces the model that we propose for the e-government ontology. This model permits not only taking into account the concepts existing in different languages (multilingualism), but also lexical relations within one language and among different languages simultaneously. This allows us to establish lexical equivalents: synonymy and translation. Our approach is based on the identification and modeling of several explicit relations in e-gov.

In our case, we will adapt French as a reference language. Translations will be suggested in Arabic and in other languages. The three representation levels in our model are: (1) concepts that model abstract meanings, (2) expressions which are specific to a given language and (3) variants of expression representing different forms that it can take.

Abstract concepts constitute the hierarchy and the semantic structure of the ontology. The expressions are not organized by a hierarchical structure or by their semantic relations. Every expression is a distinct entity in every language which can be related to concepts or to other expressions and other variants of the same expression.

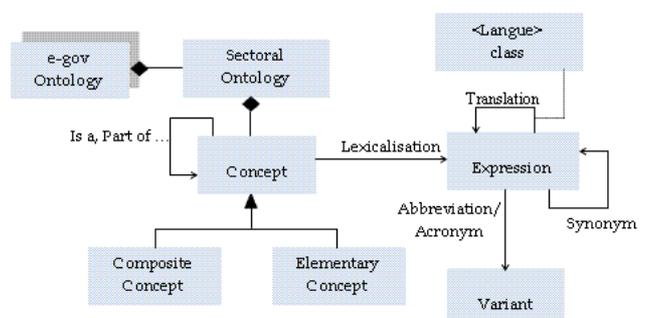

Fig. 2 The model of our ontology

The variants of expressions are not, in fact, new expressions but are variable forms of the same expression. The figure 3 presents an example representing ontology model in the customs domain along with the concept of 'Duty-free'.

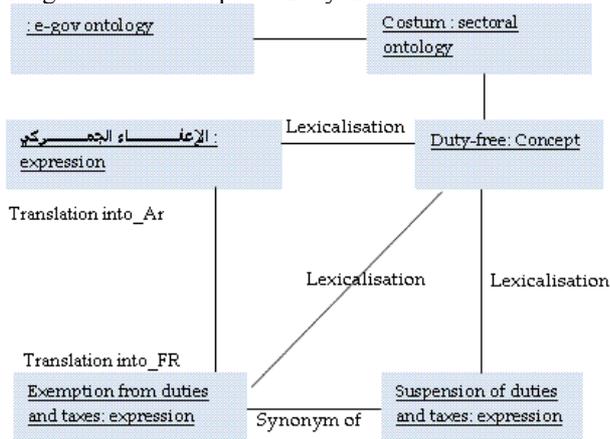

Fig. 3 A sample of the ontology model in the custom's domain.

The accomplished ontology was carried out within the collaboration between the Mohammadia School of Engineers and the Ministry of Economic and General Affairs [16].

## 4. Construction of the Ontology

4.1 Construction Method

The domain ontology construction may be linked to a number of problems. In our case, we distinguish three types of problems that are related to the implementation methodology, the application sector (domain) and the enriching of the ontology by domain and ontology experts.
The methods for ontology implementation are numerous. This is due to the differences between domains. The most known methods are: the Uschold and King's method, the SENSUS method, the On-To-Knowledge method, the Grüninger and Fox method, the KACTUS method, The Bachimont method, the METHONTOLOGY method and the Stanford University method.

The method proposed by Uschold and King is based on the experience gained during the development of the enterprise ontology. The process of this method consists of four steps: (i) identification of goals and context of the ontology, (ii) construction of the ontology, (iii) evaluation of ontology and (iv) documentation of the ontology [17].
The principle of the SENSUS method is to develop a domain ontology from greater one, SENSUS ontology [4]. This method is proposed to link the specific terms of this domain and delete the irrelevant terms. The result of this process constitutes the skeletal ontology of this new ontology.

The purpose of the method On-To-Knowledge [18] is to apply ontologies to electronically available information to improve the quality of knowledge management in large and distributed organizations. In this project, a methodology and tools for intelligent access to sources of large volume of textual information source in intra-, extra-, and internet-based environments have been developed.
The method Grüninger and Fox [19] is based on experience of project development TOVE (Toronto Virtual Enterprise Ontology project) which is an ontology in the field of modeling activities and process cases.

The KACTUS method depends on the development of applications. Therefore, each time that an application is built, the ontology representing the knowledge required for the application is built. The applications that were created are in the field of diagnosis of error in power systems. The ontology can be developed by reusing other ontologies and can also be integrated into the ontology of future applications [20].

The Bachimont method proposes to compel the user engagement by introducing a semantic standardizing of handled terms in the ontology. The normalization method follows three steps: Standardization semantic, knowledge formalization and operationalization of knowledge [21].

The METHONTOLOGY method is developed by the artificial intelligence laboratory of the Polytechnic University of Madrid. This method includes firstly the identification of the development process of the ontology, secondly the life cycle based on the development of prototypes and finally the techniques of project management, development and support activities [22].

The method is developed by Stanford University [23] has seven steps : Determination of the domain and the scope ontology, reuse of existing ontologies, listing the key terms of ontology, definition of classes and class hierarchy, definition of class properties (attributes), definition of facets of attributes, and creation of instances of classes in the hierarchy.

The choice of a method needs to take into consideration the cost and the risks throughout the process of completing the project and to ensure the quality of any result achieved. It also needs to take into account the context of specificity of the targeted ontology. It must yield efficient methodological guidelines for a better construction. In our case, we chose the method proposed by the Stanford University because it includes clear steps which are simple and easy to understand. In addition, the tool used to build the ontology in the occurrence "Protégé" is developed by the same university.

The ontology we are about to constructs is part of e-government. It has to do with sectoral ontology destined to the custom's domain which is an important pillar in Moroccan e-government. The following section presents the practical construction of this ontology.

## 4.2 Implementation of a prototype

The implementation of our ontology follows the steps of the Stanford university method such as: the delimitation of the domain and the scope of the ontology, the reuse of existing ontologies, the enumeration of important terms of the ontology, the definition of classes and their hierarchy, the definition of class properties (attributes), the definition of the attribute's facets and the creation of instance classes.

Domain and scope of the ontology: The ontology is used by the personalized search system of the customs services. The aim of this ontology is to enrich the user query during his/her search. The ontology has to meet the user requirements by taking into account the e-gov semantic.

**Reuse of existing ontologies:** Our search of existing ontologies, in the customs domain, has not produced satisfactory results. So, we have built our entire ontology. As we have already mentioned, construction of ontology is strongly related to laws and regulations of the concerned government. Consequently, it was difficult to reuse existing customs ontology for the Moroccan context.

**Enumeration of the important terms of the ontology:** The study of the customs domain allowed us to build a large list of terms. To do this, we relied on official documentation such as customs administrative documents and the online customs services.

**Definition of classes and their hierarchy:** We have identified a large number of classes from the list previously built. Each class has a name, a description, a translation into Arabic and English, synonyms and, in some cases, acronym or abbreviation.

**Definition of class properties and relations:** Each class is described by a set of properties. Each property has a name, description, type and class of membership. The relations express a connection between two concepts. Each relation has a name, a source concept, a targeted concept and cardinality.

To implement our ontology, we have chosen Protégé 4.1 for several reasons (cf. Fig. 4.). Indeed, it has an extensible architecture and provides an environment Plug-and-Play which allows having a flexible base for rapid prototyping and applications development. It can also include a general inference engine such as JESS or specific one like RACER for domain ontology. Protégé solves the problem of multilingualism in particular the Arabic language used in our ontology. The ontologies of Protégé can be exported in various formats such as RDF-Schema (RDFS), Web Ontology Language (OWL), DARPA Agent Markup Language (DAML), Ontology Inference Layer (OIL) and XML Schema. In our case, we have implemented our customs ontology by using OWL format.

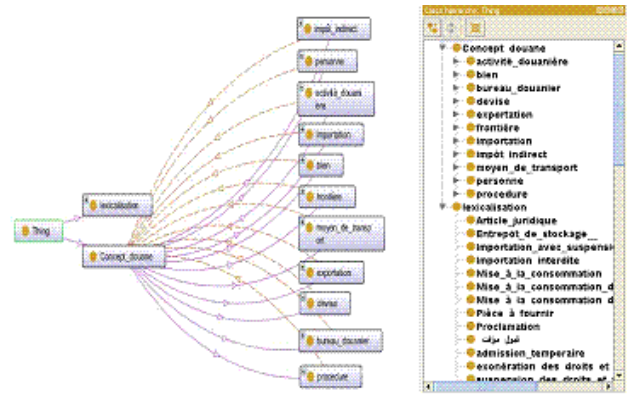

Fig. 4 Part of the Customs ontology (Protégé)

We have presented the design of our ontology and its implementation using Protégé tool. The construction methodology used was that developed by Stanford University. This choice is due to the simplicity, the scalability and the adaptability of this method to the context changing. In what follows we will present the integration of this ontology into the retrieving system.

## 5. Integration of ontology in our search platform

### 5.1 Specification of search system

Our personalized search system of e-government services is an interactive and proactive system that integrates different processes: the search, the reformulation (query), the presentation (results), the traceability (of user interactions), the recommendation and the enrichment (profile and ontology).

The search system manipulates three types of knowledge. The first is about the knowledge of the services. Its role is to describe the components of e-gov services in order to facilitate their search and to solve the heterogeneity problematic. The second relates to knowledge of the user, his interest, his preferences and his information needs. This Knowledge intervenes in the different phases of search. Finally the domain knowledge, representing a semantic layer, allows describing the concepts of multilingual terminology of a specific domain. Therefore, it lightens the ambiguity of the used terms during the formulation queries.

Our search system then integrates the user profile, domain ontology and a descriptive database of e-government services as an important component of this platform. It consists of three main functional layers: "persistence layer", "processes layer" and "communication layer" (cf. fig. 5.).

Persistence layer (low layer) represents the performed processing on the data of the search system. It concerns the storage of information in the user profile, in the ontology and in the basic descriptive e-government services.

Process layer (midle layer) includes a set of services offered by the system: personalized search services and user profile management services, the e-government services descriptive data-base and the domain ontology.

Communication layer (hight layer) is none other than the GUI. It effectuates the connection between the system and the user, and allows him/her both looking for e-government services and executing services offered by the platform. This layer permits likewise to administrate and to update the e-Government services.

## 5.2 Integration

Ontology is crucial component. It represents one of the pillars of our services search architecture (cf. Fig. 6). It is involved in the process of queries' reformulation. As, the user does not often know how to choose the appropriate words to express their information needs, the queries' reformulation generates a new query enriched and more adequate than the original one. This process is effectuated in two consecutive steps. The first consists in applying a filtering process that allows keeping the key concepts and eliminating unnecessary words. The second aims to enrich the filtered query with new terms from our ontology (cf. Fig. 7).

Augmenting the original query is a very delicate and sensitive process because of the risk of inserting incorrect information that would cause the return of result without any interest. The query's reformulation remains, actually, the most important phase in the search process. To perform better retrieving, two crucial points must be checked. The first is to achieve a good filtering and the second consists in effectuating an excellent design and a good enrichment of the ontology.

To implement our e-government services personalized search system, we used JEE technology along with three layers architecture (cf. figure 5): presentation layer, processing layer and data layer. The integration of the ontology is part of processing layer. The implementation of our architecture is technically complex, it is to be presented in at a future paper.

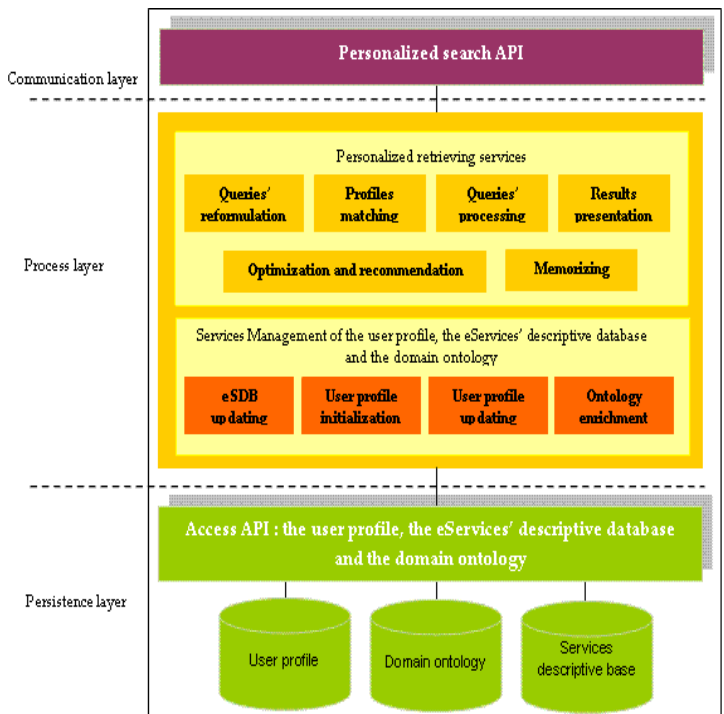

Fig. 5 Functional layers of the research system

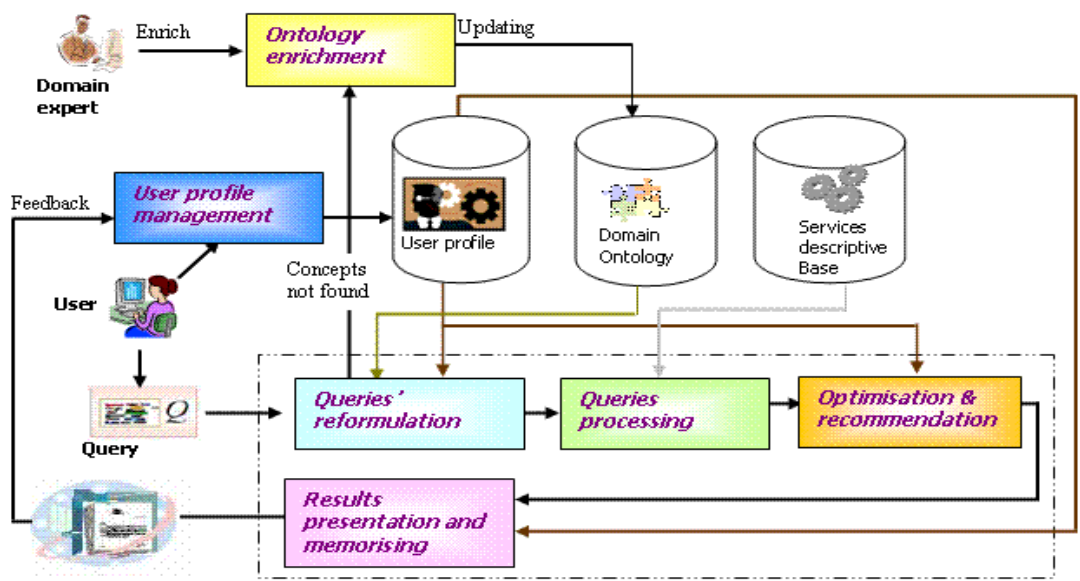

Fig. 6 E-gov services retrieval system architecture

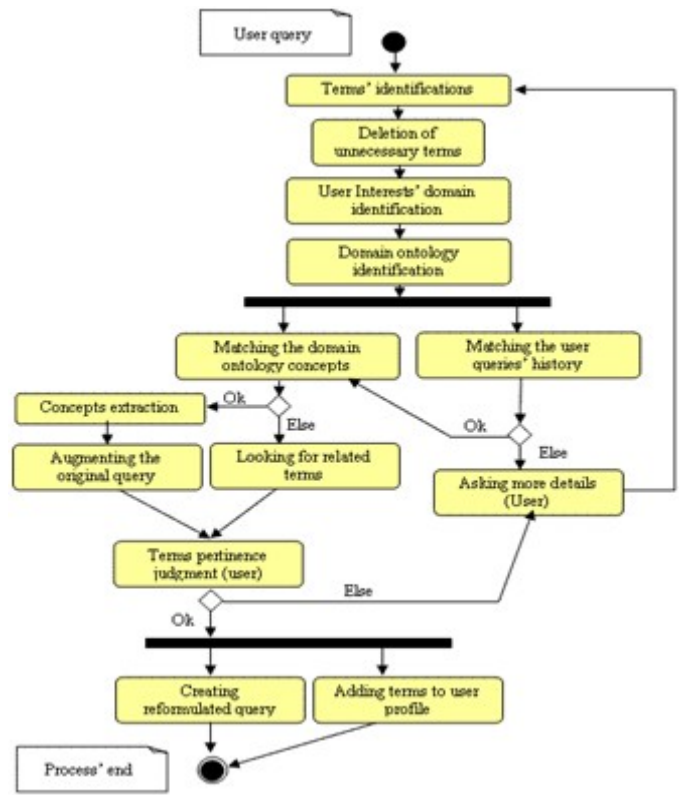

Fig. 7 Query's reformulation process

## 6. Conclusion

We have proposed, through this work, an original approach that allows an easy and personalized access to e-government services. Our approach incorporates a semantic layer as a crucial element in the retrieving services process. This consists in implementing a personalized search system that integrates an ontology of the e-gov domain in this process.

To build our ontology, we have divided the e-gov ontology into sub-ontologies (sectoral ontologies) because the terminology relating to the domain of e-government is very varied. In addition, to design this ontology, we have suggested a model which permits not only taking into account the concepts existing in different languages (multilingualism), but also lexical relations within one language and among different languages simultaneously. This ontology has been integrated into our search system. It allows the queries reformulation which generates a new enriched and more adequate query than the original one. As practical case, we have taken the customs domain.

In the perspective of the present work, we intend to develop the core of this platform and to evaluate experimentally the impact of ontology integration on the search performance.

**Hassania Ouchetto:** Engineer degree in Computer Science in 1999; DESA degree in Computing Networks and Multimedia in 2006; PhD Student in Computer Science; Chief of the Computer Science department in Ministry of Economic and General Affairs, Morocco; Ongoing research interests: e-governement, Information systems, patterns, ontology, .components

**Ouail Ouchetto** was born in Beni Mellal, Morocco. He received the Master degree in applied mathematics from the Pierre et Marie Curie, Paris, France and the Master degree in informatics from Télécom Bretagne, Brest, France. He received the PhD degree in modeling and engineering science from the University Paris Sud, Orsay, France, in 2006. He is currently a Professor with the University Aïn Chock Hassan II, Casablanca, Morocco. His research interests include electronical engineering, numerical computation, computer science and e-gov.

**Ounsa Roudiès**: Doctorate degree in Computer Science in 1989, PhD in Computer Science in 2001; Former chief of the Computer Science Department at the Mohammadia School of Engineers (EMI); Chief of Computer Science field at EMI; Professor at the Computer Science Department, EMI; Co-Editor of the etiJournal; Ongoing research interests: e-gov, IS, composition, Web services, patterns, Quality.